\documentclass[3p]{elsarticle}

\usepackage[utf8]{inputenc}
\usepackage{amssymb,bm,amsmath}
 \usepackage[colorlinks,allcolors=blue]{hyperref}
\usepackage{graphicx,subfigure}

\def\eeq{\relax}
\def\beq#1#2\eeq{\begin{equation}\label{#1}#2\end{equation}}
\def\bal#1#2\eal{\begin{align}\label{#1}#2\end{align}}
\def\bse#1#2\ese{\begin{subequations}\label{#1}#2\end{subequations}}
\def\ba{\begin{aligned}}   \def\ea{\end{aligned}}

\begin{document}
\title{Green's Function Approach to Model Vibrations of Beams with Spatiotemporally Modulated Properties}

\author{Benjamin M.~Goldsberry$^{1}$, Andrew N.~Norris$^{2}$, Samuel P.~Wallen$^{1}$, and Michael R.~Haberman$^{1,3}$}

\address{$^{1}$Applied Research Laboratories, The University of Texas at Austin, Austin Texas, 78766 USA \\
$^{2}$Mechanical and Aerospace Engineering, Rutgers University, Piscataway, New Jersey, 08854 USA\\
$^{3}$Walker Department of Mechanical Engineering, The University of Texas at Austin, Austin Texas, 78712 USA}

\begin{abstract}
The forced time harmonic response of a spatiotemporally-modulated elastic beam of finite length with light damping is derived using a novel  Green’s function approach. Closed-form solutions are found that highlight unique mode coupling effects that are induced by spatiotemporal modulation, such as split resonances that are tunable with the modulation parameters.   These effects of order unity are caused by  spatiotemporal modulation with small amplitude appropriately scaled to the magnitude of the light damping.  The scalings identified here  between the modulation amplitude, the  damping, and the inner range of frequency near the modified resonances,  translate over to more complicated and higher dimensional elastic systems. 
\end{abstract}	
%
\maketitle

\section{Introduction}\label{sec1}

Elastic and acoustic media with spatiotemporally-modulated (STM) material properties have been shown to support unique dynamical phenomena such as nonreciprocal wave propagation \cite{trainiti2016non,nassar2017non,nassar2017metabeam}, topologically protected interface states \cite{fleury2016floquet,oudich2019space}, and enhanced sound diffusion performance \cite{kang2022sound}. Prior research on STM elastic media has primarily focused on infinite domains, with performance quantified via Bloch wave analysis \cite{vila2017bloch,riva2020non}. For example, nonreciprocity is indicated by an asymmetric tilt of the band structure, which generates unidirectional band gaps and highly asymmetric propagation \cite{trainiti2016non,wallen2019nonreciprocal}. While some predictions from infinite, theoretical models may hold qualitatively for wave behavior in finite systems that have been created for experimental demonstration \cite{chen2019nonreciprocal,marconi2020experimental,attarzadeh2020experimental}, explicit models of finite, STM structures are limited. In recent works, the authors have examined damped-driven vibrations of finite, straight \cite{Goldsberry2020} and curved \cite{Goldsberry2022} Euler-Bernoulli beams with STM, revealing a complex parameter space in which strongly nonreciprocal dynamics are relatively sparse and depend on the amount of dissipation present in the system.

The Green's function, or impulse response, of a system is critical for understanding the response of a particular system under general forcing conditions. Mathematically, the Green's function of a system described by a set of partial differential equation (PDE) subjected to boundary conditions can be determined via the particular solution of the inhomogeneous PDE with an impulse forcing function that is mathematically described by Dirac delta functions \cite{feshbach2019methods}. The response of the system is then determined by convolving the impulse response (Green's function) with the desired forcing function. In linear, time-invariant systems, the impulse response can be experimentally determined using deconvolution techniques given a known broadband excitation, such as a chirp signal \cite{berkhout1980new, cushing2022design}.
Examples of the determination of the Green's function for systems with STM material properties in the literature is currently sparse due to the increased mathematical complexity of the PDEs \cite{ptitcyn2023tutorial}. However, the Green's function of STM media displays interesting properties that are not typically found in linear-time-invariant media, including providing an analytical means to determine the nonreciprocity of a physical system via the asymmetry present in the Green's function. For example, the characterization of the nonreciprocal Green's function is crucial in understanding direct measurements in dynamic media, such as underwater acoustics \cite{wapenaar2006nonreciprocal,snieder2007unified, roux2004nonreciprocal,ptitcyn2023tutorial} and robotic metamaterials \cite{brandenbourger2019non}.

The primary contribution of the present work is to establish a Green's function (GF)-based framework to study the damped-driven response of STM elastic media. To this end, we derive the GF for flexural vibrations of a finite, Euler-Bernoulli beam whose Young's modulus is modulated in space and time, and use it to investigate the dynamical behavior of the system.
The Green's function is then used to investigate the driven response of a damped finite Euler-Bernoulli beam near the resonance frequencies of the unmodulated beam, highlighting unique mode coupling effects induced by STM as a function of the modulation parameters.
The paper is structured as follows: Section \ref{sec2} defines the  GF problem in the context of a one dimensional, finite system with STM material properties and a general solution is given in terms of infinite summations and infinite systems of equations. An exact closed-form solution for the nonreciprocal Green’s function is derived in Section \ref{sec3}, where it is shown to be suitable for asymptotic expansion in a small parameter defining the modulation amplitude, and its general properties are discussed. Section \ref{sec4} examines novel and unique phenomena associated with the effects of STM on the resonance structure of the finite system. A particular scaling behavior is identified relating the material damping with the small modulation amplitude under which resonances are split and shifted by order unity when certain conditions are met. Section \ref{sec5} summarizes the work.
%
\section{The modulated system equations of motion}\label{sec2}
\subsection{Flexural waves in a spatiotemporally-modulated finite beam}
We seek an explicit expression for the flexural wave Green's function of a thin  beam of length $L$ with space- and time-dependent material properties given a time-harmonic point force located at $x_s$ with angular frequency $\omega_0$. 
For low drive frequencies, the vibrational response of the beam can be well approximated with an Euler-Bernoulli beam model.
The Green's function $g(x,t)$ therefore satisfies the following partial differential equation: 
\beq{GF}
\mathcal{L}g = \delta(x - x_\mathrm{s})e^{-i\omega_0 t}, \quad 0 \leq x \leq L,
\eeq
plus the given pinned, fixed, or free boundary conditions at each end of the beam.
The operator $\mathcal{L}$ in Eq.~\eqref{GF} is the flexural wave equation of the form
\beq{LDef}
\mathcal{L} = R_g^2\frac{\partial^2}{\partial x^2} \left(E(x,t) \frac{\partial^2}{\partial x^2}\right) + \rho(x,t)\frac{\partial^2}{\partial t^2} - Z \frac{\partial}{\partial t},
\eeq
where $R_g$ is the radius of gyration, $Z$ represents the mechanical impedance of the viscous forces present on the beam, and $E(x,t)$ and $\rho(x,t)$ are the spatiotemporally-modulated Young's modulus and density, respectively. 
The material property modulations can be decomposed into static and spatiotemporally-modulated components,
\begin{align}
E(x,t) &= E_0 + \epsilon E'(x,t), \\
\rho(x,t) &= \rho_0  + \epsilon \rho'(x,t),
\end{align}
where $\epsilon$ is the dimensionless modulation amplitude which is assumed to be a small quantity. 
Similarly, $\mathcal{L}$ can be decomposed as 
$\mathcal{L} = \mathcal{L}_0 + \epsilon \mathcal{L}'$, where
\begin{align}  \label{L0Def}
    \mathcal{L}_0 &= R_g^2E_0\frac{\partial^4}{\partial x^4} + \rho_0 \frac{\partial^2}{\partial t^2} - Z \frac{\partial}{\partial t}, \\ 
    \mathcal{L}' &= R_g^2\frac{\partial^2}{\partial x^2} \left(E'(x,t) \frac{\partial^2}{\partial x^2}\right) + \rho'(x,t)\frac{\partial^2}{\partial t^2}. \label{LprimeDef}
\end{align}
The operator $\mathcal{L}_0$ is the flexural wave equation of the unmodulated beam, and $\mathcal{L}'$ contains the material modulation terms.

\subsection{Formal solution in terms of the unmodulated modes}

We are interested in the vibrational response of the beam for drive frequencies $\omega_0$ near the resonance frequencies of the unmodulated system, which is defined as the eigenvalue problem
\bal{NonModulatedSys}
\bar{\mathcal{L}}_0(\omega_j) \psi_j &= 0, \\
\bar{\mathcal{L}}_0(\omega_j) &= R_g^2E_0\frac{\partial^4}{\partial x^4}  - \rho_0 \omega_j^2,
\eal
where $\bar{\mathcal{L}}_0(\omega_j)$ is the frequency-domain operator of Eq.~\eqref{L0Def} with no viscous forces present ($Z=0$), and $\psi_j$ is the $j$-th canonical mode shape of the beam with resonance frequency $\omega_j$.
We assume that the material property modulations in Eq.~\eqref{LDef} are periodic in time with modulation frequency $\omega_\mathrm{m}$.
Therefore, the Green's function solution to Eqs.~\eqref{GF}-\eqref{LDef} can be found in the frequency domain by expanding the variation of the Young's modulus and density as a Fourier series in time,
\begin{align}\label{eq:FourierSeresL}
    E'(x,t) &= \sum \limits_{\substack{q=-\infty\\ q \neq 0}}^\infty \hat{E}^q(x) e^{-\mathrm{i} q \omega_\mathrm{m} t}, \\
    \rho'(x,t) &= \sum \limits_{\substack{q=-\infty\\ q \neq 0}}^\infty \hat{\rho}^q(x) e^{-\mathrm{i} q \omega_\mathrm{m} t}, \label{eq:FourierSeriesRho}
\end{align}
where $\hat{E}^q(x)$ and $\hat{\rho}^q(x)$ are the spatial profiles of the Young's modulus and density, respectively, for each frequency component.
Likewise, the Green's function can also be expanded as
\beq{GFFourierSeries}
g(x,t) = \sum \limits_{p=-\infty}^\infty \hat{g}^p(x) e^{-\mathrm{i} \left(\omega_0 + p \omega_\mathrm{m}\right) t}.
\eeq
The unknown Green's function spatial profiles at each frequency $\omega_0 + p\omega_\mathrm{m}$, written in Eq.~\eqref{GFFourierSeries} as $\hat{g}^p(x)$, can be represented as a weighted sum of the mode shapes of the unmodulated beam since the eigenfunctions $\psi_j$ form a complete orthogonal set for describing the beam displacement given the specified boundary conditions \cite{Goldsberry2020}. The spatial profile for the $p^\mathrm{th}$ frequency component of the Green's function is therefore written as
\beq{GFModeExpansion}
\hat{g}^p(x) = \sum \limits_{j=1}^\infty \hat{g}_j^p \psi_j(x),
\eeq 
where $\hat{g}^p_j$ is the modal amplitude for mode $j$ at frequency $\omega_0 + p \omega_\mathrm{m}$.
Finally, using Eqs.~\eqref{GFFourierSeries}-\eqref{GFModeExpansion}, the total Green's function solution can be written by superposing mode shapes and frequency harmonics as
\begin{equation}\label{eq:GreensFcnTotalSum}
    g(x,t) = \sum \limits_{p=-\infty}^\infty \bigg(\sum \limits_{j=1}^{\infty} \hat{g}_j^p \psi_j(x)\bigg) e^{-\mathrm{i} (\omega_0 + p\omega_\mathrm{m}) t}.
\end{equation}

The equations for the unknown harmonic modal amplitudes $\hat{g}^p_n$ are found by substituting Eq.~\eqref{eq:GreensFcnTotalSum} into Eq.~\eqref{GF}, multiplying the left- and right-hand sides of Eq.~\eqref{GF} by the mode shape $\psi_k$, and integrating over $L$. The integration of the $\mathcal{L}_0$ operator terms are
\begin{equation}
    \int \limits_0^L \psi_k \mathcal{L}_0 g\, dx = \left[\rho_0 (\omega_0 + p\omega_\mathrm{m})^2 - \rho_0\omega_k^2 + iZ(\omega_0 + p \omega_\mathrm{m})\right] \hat{g}_k^p,
\end{equation}
which was simplified by utilizing the orthogonality of the Fourier series in time and the mode shapes.
The integration of $\mathcal{L}'$ requires more care. We first write the integral of the $\mathcal{L}'$ terms as
\begin{equation} \label{eq:LpDev1}
    \int \limits_0^L \psi_k \mathcal{L}' g\, dx = \int \limits_0^L \psi_k \frac{\partial^2}{\partial x^2} \left(E'(x,t) \frac{\partial^2 g}{\partial x^2}\right)\, dx + \int \limits_0^L \rho'(x,t) \psi_k \frac{\partial^2 g}{\partial t^2}\, dx.
\end{equation}
Integration by parts twice on the first integral of the right hand side of \eqref{eq:LpDev1} yields
\begin{multline} \label{eq:lpDev2}
    \int \limits_0^L \psi_k \mathcal{L}' g\, dx = \int \limits_0^L E'(x,t) \frac{\partial^2 \psi_k}{\partial x^2}\frac{\partial^2 g}{\partial x^2}\, dx 
    + \int \limits_0^L \rho'(x,t) \psi_k \frac{\partial^2 g}{\partial t^2}\, dx \\
    + \left[\psi_k\frac{\partial}{\partial x}\left(E'(x,t)\frac{\partial^2 g}{\partial x^2}\right) - \frac{\partial \psi_k}{\partial x}E'(x,t)\frac{\partial^2 g}{\partial x^2}\right]\bigg|_0^L.
\end{multline}
Note that the boundary terms in Eq.~\eqref{eq:lpDev2} are zero since the mode shapes and Green's function satisfy either pinned, clamped, or free boundary conditions at each end of the beam.
Equation \eqref{eq:lpDev2} is transformed to the frequency domain by substituting Eqs.~\eqref{eq:FourierSeresL}, \eqref{eq:FourierSeriesRho}, and \eqref{eq:GreensFcnTotalSum} into Eq.~\eqref{eq:lpDev2} and using the orthogonality of the mode shapes and Fourier series in time,
\begin{equation}
    \int \limits_0^L \psi_k \mathcal{L}' g\, dx = \sum \limits_{s} \sum \limits_j \hat{g}^s_j\left[\int \limits_0^L \hat{E}^{p-s}(x)\frac{\partial^2 \psi_k}{\partial x^2}\frac{\partial^2 \psi_j}{\partial x^2}\, dx   + (\omega_0 + s \omega_\mathrm{m})^2\int \limits_0^L \hat{\rho}^{p-s}(x) \psi_k \psi_j\, dx\right].
\end{equation}
Truncation of the above system for $N$ spatial modes and $2P+1$ modulation harmonics produces a  system of equations for the modal amplitudes, which can be written as a matrix-vector system
\begin{equation}\label{eq:GreensFcnMatVec}
    \mathbf{\Omega}_p\hat{\bm{g}}_p =  \bm{\psi}_\mathrm{s} + \epsilon \sum \limits_{s=-P}^P \left(\mathbf{K}_{p-s} + (\omega_0 + s\omega_\mathrm{m})\mathbf{R}_{p-s}\right)\hat{\bm{g}}_s, \quad p \in [-P,P]
\end{equation}
where $\hat{\bm{g}}_p = [\hat{g}_1^p, \hat{g}_2^p, ..., \hat{g}_N^p]^\mathrm{T}$ represents the vector of unknown amplitudes for each mode shape for the $p^\mathrm{th}$ modulation harmonic, $\bm{\psi}_\mathrm{s} = [\psi_1(x_\mathrm{s}), \psi_2(x_\mathrm{s}), ..., \psi_N(x_\mathrm{s})]^\mathrm{T}$  represent the mode shapes evaluated at the source $x_\mathrm{s}$, and

\begin{align} \label{eq:OmegaMat}
    \left(\Omega_p\right)_{jk} &= \left[\rho_0 (\omega_0 + p\omega_\mathrm{m})^2 - \rho_0\omega_k^2 + iZ(\omega_0 + p \omega_\mathrm{m})\right] \delta_{jk}, \\
    \left(K_{p-s}\right)_{jk} &= \int \limits_0^L \hat{E}^{p-s}(x)\frac{\partial^2 \psi_k}{\partial x^2}\frac{\partial^2 \psi_j}{\partial x^2}\, dx, \label{eq:KMat}\\
    \left(R_{p-s}\right)_{jk} &= \int \limits_0^L \hat{\rho}^{p-s}(x) \psi_k \psi_j\, dx. \label{eq:RMat}
\end{align}

For simplicity, we assume that only the Young's modulus is spatiotemporally modulated ($\rho'(\bm{x},t) = 0$), and that the modulation function $E'$ of  Eq.~\eqref{eq:FourierSeresL} contains only one Fourier component.  The modulation is then a single traveling wave, i.e.~$E'(x,t) = E_0\cos(2\pi x/L - \omega_\mathrm{m} t)$, implying 
$ \hat{E}^q =  \hat{E}^1\delta_{q,1} + \hat{E}^{1*}\delta_{q,-1} $
where $\hat{E}^1 = 0.5E_0\,e^{\mathrm{i}2\pi x/L}$. 
Therefore, the only $K$-matrices of Eq.~\eqref{eq:GreensFcnMatVec} are $\mathbf{K}_1 \equiv \mathbf{K}$, and $\mathbf{K}_{-1} = \mathbf{K}^*$, where  $*$ denotes complex-conjugation.
Closed-form solutions to Eq.~\eqref{eq:GreensFcnMatVec}  are difficult to find due to the number of modes and harmonics that need to be retained in order to compute an accurate solution, but can be computed numerically \cite{Goldsberry2020}. 
However, only a small number of modes are needed when the drive frequency is near a resonance of the beam, as the beam displacement is mostly described by that mode shape.
In this work, we are interested in investigating the drive frequency component of the Green's function, $\hat{\bm{g}}_0$, though a similar analysis can also be applied to the other frequency components. Closed-form expressions can be derived by taking advantage of the explicit scaling of the modulation in Eq.~\eqref{eq:GreensFcnMatVec} as will be demonstrated in Sec.~\ref{sec3}.

\section{Exact closed-form solution}\label{sec3}
Solutions to \eqref{eq:GreensFcnMatVec} for the drive frequency component $\hat{\bm{g}}_0$ are now sought for finite values of the harmonic number $P$. We first consider the simple cases $P=1$ and $P=2$, and thereby deduce a recursion method of solution for arbitrary $P$. 

\subsection{Solutions for  arbitrary $P$}\label{sec3a}
We begin with the simplest case of  $P=1$, for which Eq.~\eqref{eq:GreensFcnMatVec} can be written as
\bse{1=2}
\begin{align}
  \mathbf{\Omega}_{-1}\hat{\bm{g}}_{-1} &= \epsilon \mathbf{K}^*\hat{\bm{g}}_0, \label{eq:Gm1}\\
  \mathbf{\Omega}_0\hat{\bm{g}}_0 &= \bm{\psi}_\mathrm{s} + \epsilon\mathbf{K}^* \hat{\bm{g}}_1 + \epsilon\mathbf{K}\hat{\bm{g}}_{-1}, \label{eq:G0}\\
  \mathbf{\Omega}_1 \hat{\bm{g}}_1 &= \epsilon \mathbf{K} \hat{\bm{g}}_0. \label{eq:Gp1}
\end{align}
\ese
Equations \eqref{eq:Gm1} and \eqref{eq:Gp1} are substituted into Eq.~\eqref{eq:G0} to yield the following expression for $\hat{\bm{g}}_0$:
\begin{equation} \label{eq:G0Implct}
  \mathbf{\Omega}_0 \hat{\bm{g}}_0 = \bm{\psi}_\mathrm{s} + \epsilon^2\mathbf{\Gamma}_1 \hat{\bm{g}}_0,
\end{equation}
where 
\beq{1=4}
\mathbf{\Gamma}_1 = \mathbf{K}^*\overline{\mathbf{\Omega}}_1\mathbf{K} + \mathbf{K}\overline{\mathbf{\Omega}}_{-1}\mathbf{K}^* 
\ \ \text{and }\ \ \overline{\mathbf{\Omega}}_s = (\mathbf{\Omega}_s)^{-1}.  
\eeq

For the case of $P=2$, Eq.~\eqref{eq:GreensFcnMatVec} can be written as
\bse{2-1}
\begin{align}
  \mathbf{\Omega}_{-2}\hat{\bm{g}}_{-2} &= \epsilon \mathbf{K}^*\hat{\bm{g}}_{-1}, \label{eq:Gm2b}\\
  \mathbf{\Omega}_{-1}\hat{\bm{g}}_{-1} &= \epsilon\mathbf{K}^*\hat{\bm{g}}_0 + \epsilon\mathbf{K}\hat{\bm{g}}_{-2}, \label{eq:Gm1b}\\
  \mathbf{\Omega}_0\hat{\bm{g}}_0 &= \bm{\psi}_\mathrm{s} + \epsilon\mathbf{K}^* \hat{\bm{g}}_1 + \epsilon\mathbf{K}\hat{\bm{g}}_{-1},\label{eq:G0b}\\
  \mathbf{\Omega}_1 \hat{\bm{g}}_1 &= \epsilon \mathbf{K} \hat{\bm{g}}_0 + \epsilon\mathbf{K}^*\hat{\bm{g}}_2, \label{eq:Gp1b}\\
  \mathbf{\Omega}_2\hat{\bm{g}}_2 &= \epsilon \mathbf{K}\hat{\bm{g}}_1. \label{eq:Gp2b}
\end{align}
\ese
Eliminating $\hat{\bm{g}}_{-1}$ and $\hat{\bm{g}}_1$ using  Eqs.~\eqref{eq:Gm1b} and \eqref{eq:Gp1b}, the remaining three equations for $\hat{\bm{g}}_{-2}$, $\hat{\bm{g}}_0$ and $\hat{\bm{g}}_2$ become
\bse{2=2}
\begin{align}
  \mathbf{\Omega}_{-2}\hat{\bm{g}}_{-2} &= \epsilon^2 \mathbf{K}^* \overline{\mathbf{\Omega}}_{-1}\mathbf{K}^* \hat{\bm{g}}_0 + \epsilon^2 \mathbf{K}^* \overline{\mathbf{\Omega}}_{-1}\mathbf{K}\hat{\bm{g}}_{-2}, \label{eq:Gm2c}\\
   \mathbf{\Omega}_0 \hat{\bm{g}}_0 &= \bm{\psi}_\mathrm{s} + \epsilon^2 \mathbf{\Gamma}_1 \hat{\bm{g}}_0 +  \epsilon^2\mathbf{K}\overline{\mathbf{\Omega}}_{-1}\mathbf{K}\hat{\bm{g}}_{-2} +
   \epsilon^2\mathbf{K}^*\overline{\mathbf{\Omega}}_1 \mathbf{K}^*\hat{\bm{g}}_2, \label{eq:G02Harm}
  \\
  \mathbf{\Omega}_{2}\hat{\bm{g}}_2 &= \epsilon^2 \mathbf{K} \overline{\mathbf{\Omega}}_{1}\mathbf{K} \hat{\bm{g}}_0 + \epsilon^2 \mathbf{K} \overline{\mathbf{\Omega}}_{1}\mathbf{K}^*\hat{\bm{g}}_{2},\label{eq:Gp2c}
\end{align}
\ese
where $ \mathbf{\Gamma}_1$ is defined in Eq.\ \eqref{1=4}.
Finally, we substitute Eq.~\eqref{eq:Gm2c} and \eqref{eq:Gp2c} into Eq.~\eqref{eq:G02Harm} to yield an  equation for $\hat{\bm{g}}_0$:
\beq{-2}
  \mathbf{\Omega}_0 \hat{\bm{g}}_0 = \bm{\psi}_\mathrm{s} + \epsilon^2  \mathbf{\Gamma}_2  \hat{\bm{g}}_0
\eeq
where 
\beq{1=5}
\mathbf{\Gamma}_2= \mathbf{K}^* \big( \mathbf{\Omega}_{1}-\epsilon^2  \mathbf{K}^*\overline{\mathbf{\Omega}}_{2}\mathbf{K}    \big)^{-1} \mathbf{K}
+
\mathbf{K} \big( \mathbf{\Omega}_{-1}-\epsilon^2  \mathbf{K}\overline{\mathbf{\Omega}}_{-2}\mathbf{K}^*    \big)^{-1} \mathbf{K}^*.
 \eeq
 Note that Eqs.~\eqref{eq:G0Implct} and \eqref{-2} have identical mathematical forms, and the matrix $\mathbf{\Gamma}$ can be generated with a recursive pattern which is now generalized for arbitrary $P$.

 Thus, the solution for 
  the case of arbitrary $P$  is that  $\hat{\bm{g}}_0$
satisfies the following  equation 
\beq{-21}
  \mathbf{\Omega}_0 \hat{\bm{g}}_0 = \bm{\psi}_\mathrm{s} + \epsilon^2  \mathbf{\Gamma}_P  \hat{\bm{g}}_0,
\eeq
with $\mathbf{\Gamma}_P$ defined by 
\bal{2=5}
&\mathbf{\Gamma}_P= \mathbf{X}_{-1}+\mathbf{X}_1  \ \ \text{where}    \notag 
\\ 
&
\ba
 \mathbf{X}_k  &= \mathbf{K}^* \big( \mathbf{\Omega}_{k}-\epsilon^2  
  \mathbf{X}_{k+1}   \big)^{-1}\mathbf{K}, 
 \\
  \mathbf{X}_{-k}  &= \mathbf{K} \big( \mathbf{\Omega}_{-k}-\epsilon^2  
  \mathbf{X}_{-k-1}   \big)^{-1}\mathbf{K}^* , 
\ea  \bigg \}  \ \ k=P-1, P-2, \ldots, 2,1 , 
    \\
   &\mathbf{X}_P  =\mathbf{K}^* \overline{\mathbf{\Omega}}_{P}\mathbf{K},  
   \quad
   \mathbf{X}_{-P} =\mathbf{K} \overline{\mathbf{\Omega}}_{-P}\mathbf{K}^* .
   \notag
\eal
The expression  \eqref{2=5} reproduces $\mathbf{\Gamma}_1$ and $\mathbf{\Gamma}_2$
of Eqs.\ \eqref{1=4} and \eqref{1=5} for $P=1$ and $P=2$, respectively. The closed form solution of the recursion relations in \eqref{2=5} can be written out, although the expressions become lengthy. 
For instance, if $P=4$, 
\bal{1=6}
\mathbf{\Gamma}_4&= 
 \mathbf{K} \bigg( \mathbf{\Omega}_{-1}-\epsilon^2 \mathbf{K}
\Big( \mathbf{\Omega}_{-2}-\epsilon^2 
\mathbf{K}\big( \mathbf{\Omega}_{-3}-\epsilon^2 
\mathbf{K}\overline{\mathbf{\Omega}}_{-4}\mathbf{K}^* \big)^{-1} \mathbf{K}^*    \Big)^{-1}\mathbf{K}^* \bigg)^{-1} \mathbf{K}^* 
\notag \\ &
+
\mathbf{K}^* \bigg( \mathbf{\Omega}_{1}-\epsilon^2 \mathbf{K}^* 
\Big( \mathbf{\Omega}_{2}-\epsilon^2 
\mathbf{K}^*\big( \mathbf{\Omega}_{3}-\epsilon^2 
\mathbf{K}^*\overline{\mathbf{\Omega}}_{4}\mathbf{K} \big)^{-1} \mathbf{K}    \Big)^{-1}\mathbf{K} \bigg)^{-1} \mathbf{K} .
 \eal

\subsection{Properties of the Green's function}
The solution of Eq.\ \eqref{-21} is 
\begin{equation} \label{eq:g0Explicit}
  \hat{\bm{g}}_0 = \big( {\mathbf{\Omega}}_0 -\epsilon^2 \mathbf{\Gamma}_P\big)^{-1} \bm{\psi}_\mathrm{s}.
\end{equation}
which after substitution  into Eq.~\eqref{eq:GreensFcnTotalSum} for $p=0$ gives 
\begin{equation} \label{eq:GreensFcnSoln}
  g(x,t|x_\mathrm{s}) = \bm{\psi}^\mathrm{T}
           \big( {\mathbf{\Omega}}_0 -\epsilon^2 \mathbf{\Gamma}_P\big)^{-1} \bm{\psi}_\mathrm{s},
\end{equation}
where  $\bm{\psi} = [\psi_1(x), \psi_2(x), ..., \psi_N(x)]^\mathrm{T}$. 
The classical reciprocal solution of the unmodulated beam subject to the boundary conditions corresponds to 
$\epsilon= 0$, i.e.\ 
$ g(x,t|x_\mathrm{s}) = \bm{\psi}^\mathrm{T} \overline{\mathbf{\Omega}}_0   \bm{\psi}_\mathrm{s}$, 
which is reciprocal because $\mathbf{\Omega}_0$ is symmetric.
Off-diagonal terms in $\mathbf{\Gamma}_P$ indicate that the spatiotemporal modulation of the material properties can induce mode coupling. Further, if $\mathbf{\Gamma}_P$ is not symmetric, then the Green's function is nonreciprocal. Our initial inspection of the symmetry properties of $\mathbf{\Gamma}_P$ indeed show that the matrix is not symmetric in general if the loss is non-zero $(Z>0)$. This observation is consistent with observations in earlier works by the authors which showed that nonreciprocal vibrational motion in finite Euler-Bernoulli beams is only possible when losses are present \cite{Goldsberry2020}. 

The  solution $\hat{\bm{g}}_0 $ in Eq.~\eqref{eq:g0Explicit} can be expressed as a series by 
 noting $\big( {\mathbf{\Omega}}_0 -\epsilon^2 \mathbf{\Gamma}_P\big)^{-1}
= \big( {\mathbf{I}}  -\epsilon^2 \overline{\mathbf{\Omega}}_0\mathbf{\Gamma}_P\big)^{-1}
\overline{\mathbf{\Omega}}_0$ and using the standard expansion  $(1-z)^{-1} \approx 1 + z + z^2 + \cdot\cdot\cdot$. Therefore, the Green's function at the source frequency $\omega_0$ is given by 
\beq{-1}
\hat{\bm{g}}_0 =  {\mathbf{\Omega}}_0 \bm{\psi}_\mathrm{s} +\sum_{k=1}^\infty \epsilon^{2k} 
\big( \overline{\mathbf{\Omega}}_0\mathbf{\Gamma}_P \big)^k\,  \bm{\psi}_\mathrm{s} . 
\eeq
In order to express the above as an expansion in powers of $\epsilon^2$ it is necessary to re-express $\mathbf{\Gamma}_P$ as such a series, which becomes more complicated as $P$ increases. 
 For example, truncating at two terms ($k=1$) yields
\beq{-45}
  \hat{\bm{g}}_0^2 = \overline{\mathbf{\Omega}}_0\bm{\psi}_\mathrm{s} + \epsilon^2\overline{\mathbf{\Omega}}_0 \mathbf{\Gamma}_1 \overline{\mathbf{\Omega}}_0 \bm{\psi}_\mathrm{s}  
\eeq
which only involves $\mathbf{\Gamma}_1$ of Eq.\ \eqref{1=4}.  Truncating at three terms 
($k=2$) requires expanding $\mathbf{\Gamma}_2$, with the result 
 \bal{-46}
  \hat{\bm{g}}_0^3 =& \overline{\mathbf{\Omega}}_0\bm{\psi}_\mathrm{s} + \epsilon^2\overline{\mathbf{\Omega}}_0 \mathbf{\Gamma}_1 \overline{\mathbf{\Omega}}_0 \bm{\psi}_\mathrm{s} + \epsilon^4 \overline{\mathbf{\Omega}}_0 \big(  \mathbf{\Gamma}_1 \overline{\mathbf{\Omega}}_0 \mathbf{\Gamma}_1 
  \notag \\ 
  &+\mathbf{K}^*\overline{\mathbf{\Omega}}_1 \mathbf{K}^*  \overline{\mathbf{\Omega}}_2\mathbf{K}\overline{\mathbf{\Omega}}_{1}\mathbf{K} + \mathbf{K}\overline{\mathbf{\Omega}}_{-1}\mathbf{K} \overline{\mathbf{\Omega}}_{-2} \mathbf{K}^* \overline{\mathbf{\Omega}}_{-1}\mathbf{K}^*
  \big) \overline{\mathbf{\Omega}}_0 \bm{\psi}_\mathrm{s}.
\eal
This implies that the infinite series expansion \eqref{-1} is only valid to  $\mathcal{O}(\epsilon^{2P})$ and therefore the summation should be truncated at $k=P$.  

 \subsection{Summary: Leading order asymptotic solution}
 
The above results have  shown that the exact solution may  be formally represented by  an asymptotic expansion in the small parameter $\epsilon$.  We do not attempt to justify the range of applicability (i.e.\ convergence) of this expansion, but for the remainder of the paper focus on the leading order corrections to the unmodulated Green's function.  In that regard, of all the results from  Section \ref{sec3a} perhaps the most important is that $\mathbf{\Gamma}_P = \mathbf{\Gamma}_1 +$$\mathcal{O}(\epsilon^2)$.  Therefore the leading order asymptotic solution for the modulated Green's function follows from  Eq.\  \eqref{eq:GreensFcnSoln} as 
 \beq{9=6}
  g(x,t|x_\mathrm{s}) \approx  \bm{\psi}^\mathrm{T}
           \big( {\mathbf{\Omega}}_0 -\epsilon^2 \mathbf{\Gamma}_1\big)^{-1} \bm{\psi}_\mathrm{s} .
\eeq
Generally, the error incurred in replacing $\mathbf{\Gamma}_P \to  \mathbf{\Gamma}_1$ is       $\mathcal{O}(\epsilon^4)$, although it can be $\mathcal{O}(\epsilon^2)$ in some circumstances involving resonances.  However, in the latter case, as we will see below in Section \ref{sec4}, the term $\epsilon^2 \mathbf{\Gamma}_1$ gives rise to an order unity effect and hence provides the leading order modulated correction. 

\section{Resonance in the presence of modulation}\label{sec4}
\subsection{Flexural wave mode shapes}
We now investigate the Green's function for the specific cases of pinned-pinned, and clamped-free boundary conditions. The boundary conditions for the pinned-pinned case are $g(0,t) = g(L,t) = 0$ and $\partial^2 g(0,t)/\partial x^2 = \partial^2 g(L,t)/\partial x^2 = 0$, which, from Eq.~\eqref{NonModulatedSys}, yields the mode shapes
\begin{equation}
    \psi_j(x) = \sin(j \pi x/L),
\end{equation}
with resonance frequencies 
\begin{equation}\label{eq:eigPP}
   \omega_j = j^2 c_0 R_g/L^2. 
\end{equation}
The boundary conditions for the clamped-free case are $g(0,t) = \partial g(0,t)/ \partial x = 0$ and $\partial^2 g(L,t)/\partial x^2 = \partial^3 g(L,t)/ \partial x^3 = 0$, which yields the mode shapes
\begin{equation}
    \psi_j(x) = \sin(\beta_j x) - \mathrm{sinh}(\beta_j x) - \frac{\sin(\beta_jL) + \mathrm{sinh}(\beta_jL)}{\cos(\beta_j L) + \mathrm{cosh}(\beta_j L)} \left(\cos(\beta_j x) - \cosh(\beta_j x)\right),
\end{equation}
where $\beta_j$ are the roots of the transcendental equation $\cos(\beta L)\cosh(\beta L) = -1$, and the resonance frequencies are 
\begin{equation}\label{eq:eigCF}
    \omega_j = (\beta_jL)^2 c_0 R_g/L^2.
\end{equation}

 \subsection{Mode-commensurate modulation with light damping  }
We assume that the damping is {\em light} in the sense that $Z = \epsilon^2 z_0 $
for positive $z_0 $ of order unity. 
The unmodulated Green's function is then, from Eq.\ \eqref{eq:GreensFcnSoln} with 
$\mathbf{\Gamma}_P =0$, 
\beq{7=2}
  g_0(x,t|x_\mathrm{s}) \equiv \bm{\psi}^\mathrm{T}
           \overline{\mathbf{\Omega}}_0  \bm{\psi}_\mathrm{s}
           = \sum_k 
           \frac{\psi_k(x)\psi_k(x_s) }{\omega_0^2 - \omega_k^2 + \mathrm{i}\epsilon^2 z_0  \omega_0} .
\eeq
The  mode $k$ has a classic Lorentzian resonance at its natural frequency
 $\omega_k$, with maximum amplitude of order $\epsilon^{-2}$, i.e.\ what is called  a high Q resonance in the vibrations community. 
 
 \begin{figure}[t!] 
\centering
\includegraphics[width=\textwidth]{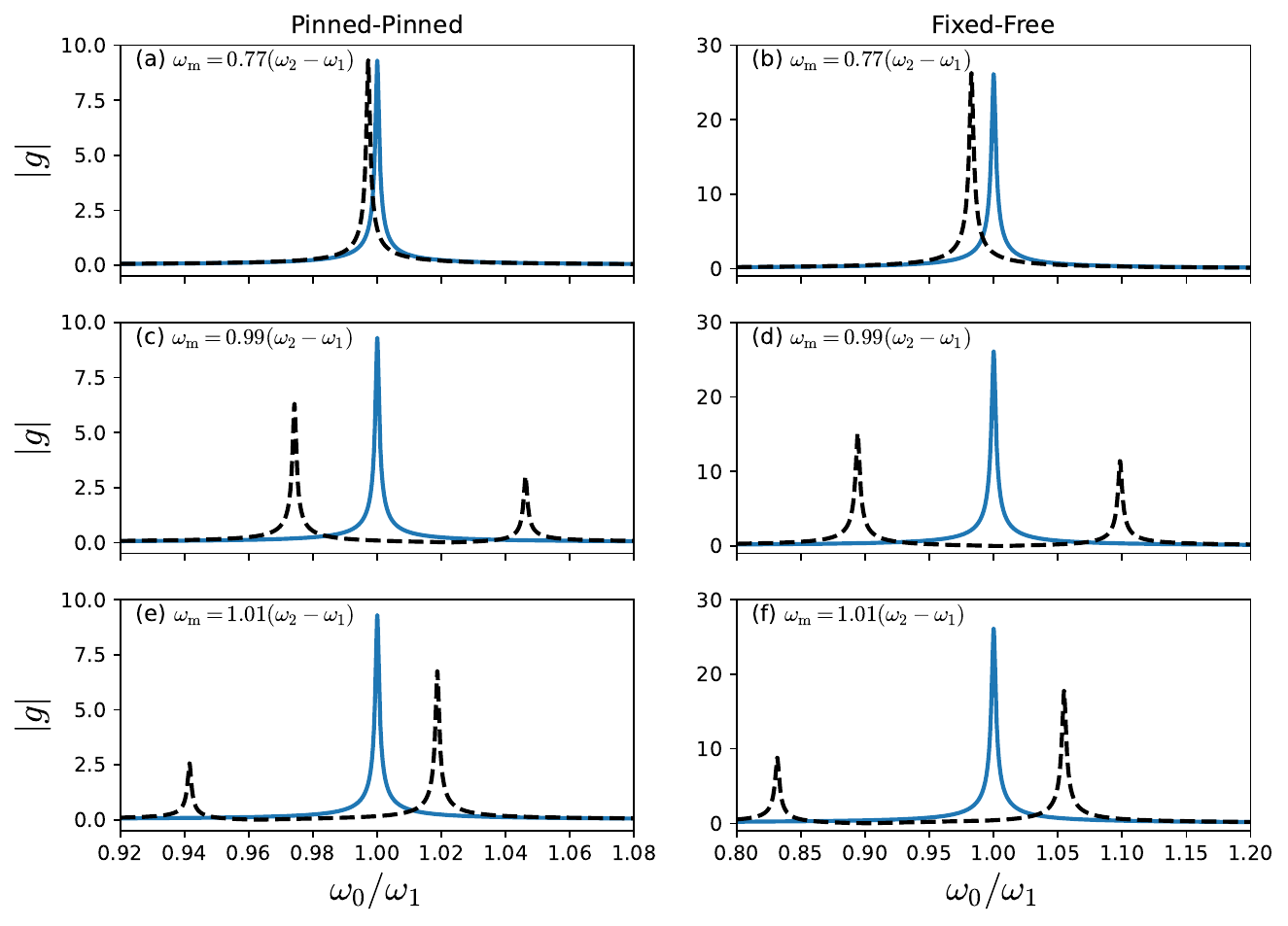}
\caption{Plots of the magnitude of the Green's function as a function of normalized frequency near the first resonance frequency for two different cases of end conditions: pinned-pinned (panels (a), (c), and (e)) and fixed-free (panels (b), (d), and (f)) and with modulation frequency $\omega_\mathrm{m}=0.77\ (\omega_2-\omega_1)$ (panels a and b), $\omega_\mathrm{m}=0.99(\omega_2-\omega_1)$ (panels c and d) and $ \omega_\mathrm{m}=1.01(\omega_2-\omega_1)$ (panels (e) and (f)). Here  $x = 0.43$, $x_s=0.61$, $L=1$, $\epsilon =0.1$ and $Z=0.01$ ($z_0=1$, see Eq.\ \eqref{7=2}).  The solid blue curves are the unmodulated response from Eq.\ \eqref{7=2}, while the dashed black curves are the modulated Green's function according to Eq.\ \eqref{9=6}.}
\label{fig1}
\end{figure}

\begin{figure}[t] 
\centering
\includegraphics[width=\textwidth]{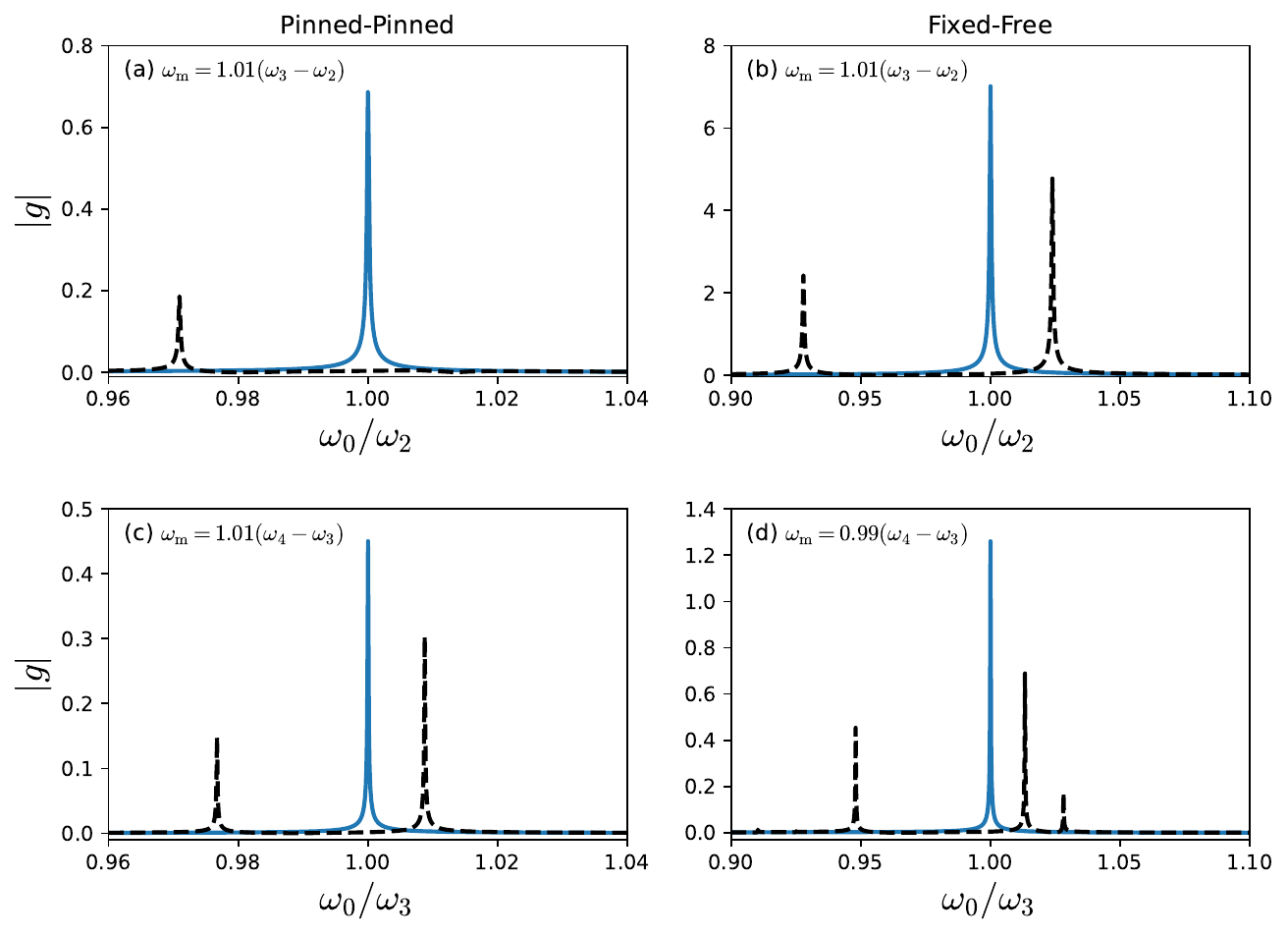}
\caption{The Green's function magnitude as a function of normalized frequency $\omega_2$ and $\omega_3$ for pinned-pinned (panels (a) and (c)) and fixed-free (panels (b) and (d)) end conditions subject to a modulation frequency of $\omega_m = 1.01(\omega_3 - \omega_2)$ (panels (a) and (b)) and $\omega_m = 1.01(\omega_4 - \omega_3)$ (panels (c) and (d)) . 
The values of the other parameters are as in Fig.\ \ref{fig1} ($x = 0.43$, $x_s=0.61$, $L=1$, $\epsilon =0.1$,  $Z=0.01$).  The dashed black curves are the modulated Green's function of Eq.\ \eqref{9=6}
and the solid blue curves show the  unmodulated response of Eq.\ \eqref{7=2}.  }
\label{fig2}
\end{figure} 
Figure \ref{fig1} shows comparisons of the unmodulated \eqref{7=2} and the modulated \eqref{9=6} responses for pinned-pinned and clamped-free end conditions near the first modal resonance $\omega_1$. 
For both cases, Eq.~\eqref{eq:g0Explicit} is numerically solved for $N=15$ modes and $P=4$ frequency harmonics to ensure convergence of the solution. 
The asymptotic parameter is  $\epsilon =0.1$, and it is evident from Figs.\ \ref{fig1}(a) and \ref{fig1}(b) that at a typical modulation frequency, here $\omega_m=  0.77(\omega_2 - \omega_1)$, that the response changes by a small amount, of order $\epsilon^2$, as expected from Eq.\ \eqref{9=6}. 
By contrast, Figs.\ \ref{fig1}(c) through  \ref{fig1}(f) show the response of the pinned-pinned and fixed-free beams for modulation frequencies close to the \textit{difference} of the first two unmodulated resonance frequencies; for  Figs.\ \ref{fig1}(c) and \ref{fig1}(d) the modulation is at a frequency slightly below the difference of the first two resonance frequencies and for Figs.\ \ref{fig1}(e) and \ref{fig1}(f) it is slightly above.  Several effects are evident: first that the resonance is split, with two resonances of different amplitudes, and secondly that the effect itself is very sensitive to the value of the modulation frequency relative to the unmodulated resonance frequency. 
Similar effects are observed when the drive frequency is close to the resonance frequency of higher modes.  Figure \ref{fig2} shows how the response for drive frequencies near the second and third modal frequencies of a pinned-pinned and fixed-free beam are altered when the modulation frequency is $\omega_m = 1.01 (\omega_3 - \omega_2)$ for drive frequencies near $\omega_2$ and $\omega_\mathrm{m} = 1.01(\omega_4 - \omega_3)$ for drive frequencies near $\omega_3$. 
\begin{figure}[t] 
\centering
\includegraphics[width=\textwidth]{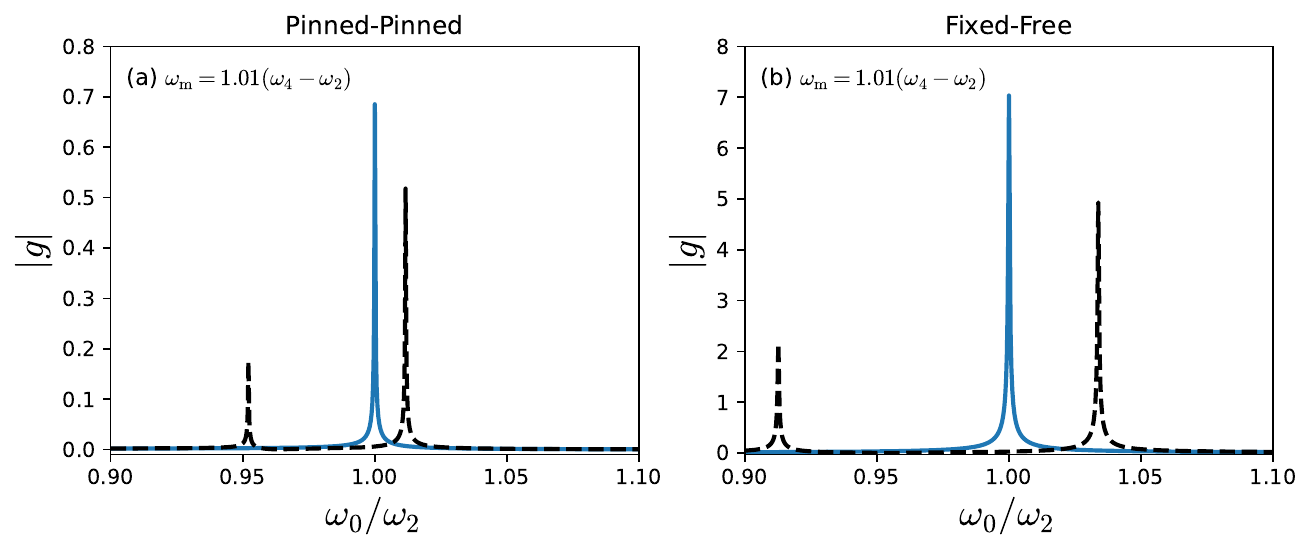}
\caption{The magnitude of the Green's function as a function of normalized frequency near the second  resonance frequency of  systems with (a) pinned-pinned and (b) fixed-free end conditions subject to a modulation frequency $\omega_m = 1.01(\omega_4-\omega_2)$. The values of the other parameters are as in Fig.\ \ref{fig1} ($x = 0.43$, $x_s=0.61$, $L=1$, $\epsilon =0.1$,  $Z=0.01$).   The dashed black curves are the modulated Green's function of Eq.\ \eqref{9=6}
and the solid blue curves show the  unmodulated response of Eq.\ \eqref{7=2}.  }
\label{fig3}
\end{figure} 

The phenomenon observed in the examples of Figs.\ \ref{fig1} and \ref{fig2}  causes a change in the resonance structure of order unity using a system modulation of order $\epsilon$.  At the same time, the expected effect at {\em most} values of the drive frequency is to shift the resonance frequency by $\mathcal{O}(\epsilon^2)$,  according to Eq.\ \eqref{9=6}.  The critical requirement for the order unity effect is that the modulation frequency is close to the {\em difference} of two unmodulated resonance frequencies.  
A modulation frequency $\omega_m \approx (\omega_{n+1} - \omega_n)$ therefore couples nearest neighbor modes to give the observed effect. 
Similarly, as Fig.\ \ref{fig3} illustrates, a modulation frequency $\omega_m \approx (\omega_{n+2} - \omega_n)$ couples next-to-nearest neighbor modes.
We next explain the mode-commensurate modulation effect in terms of a leading-order asymptotic analysis that is valid near the resonances.

 \subsection{Asymptotic analysis near the critical modulation frequencies}
Assume that the sum of the drive frequency and the modulation frequency is close to a natural frequency, specifically that of mode $\psi_r$ with frequency $\omega_r$.   We rescale the frequency in terms of a real parameter $q$ such that 
\beq{7=3}
\omega_0 + \omega_\mathrm{m} = \omega_r + \frac q2 \epsilon^2 . 
\eeq
 It follows from Eqs.~\eqref{eq:OmegaMat} and \eqref{7=3} that the elements of $\mathbf{\Omega}_1$ are
 \beq{3=71}
    \left(\mathbf{\Omega}_1\right)_{kl} = 
    \begin{cases}
    \big( \omega_r^2 - \omega_k^2 \big) \delta_{kl}  + \mathcal{O}(\epsilon^2), & k\ne r,
    \\
\epsilon^2 \omega_r (q +  \mathrm{i}  z_0 ) 
    \delta_{rl}  + \mathcal{O}(\epsilon^4), & k=r.  
    \end{cases}
\eeq
 Hence, according to the definition of $\mathbf{\Gamma}_1$ in Eq.\ 
 \eqref{1=4}, 
  \beq{1=41}
\big(\mathbf{\Gamma}_1 \big)_{kl} = 
\frac{ \epsilon^{-2}K^*_{kr}K_{rl} } {\omega_r (q +  \mathrm{i}  z_0 ) }
+  \mathcal{O}(1),
\eeq
 and, referring to the solution of Eq.\ \eqref{9=6}, 
\beq{7=5}
 \big( {\mathbf{\Omega}}_0 -\epsilon^2 \mathbf{\Gamma}_1\big)_{kl}
 =  ({\mathbf{\Omega}}_0)_{kl}
 - \frac{ K^*_{kr}K_{rl} } {\omega_r (q +  \mathrm{i}  z_0 ) } 
 + \mathcal{O}(\epsilon^2). 
 \eeq
 It follows from this result and Eq.\ \eqref{9=6} for the Green's function in the presence of modulation that the latter differs by order unity from the unmodulated Green's function \eqref{7=2}.  Note that only the elements of $\mathbf{K}$ involving mode $r$ are involved through the symmetric positive definite matrix with elements $ K^*_{kr}K_{rl}$.  Further, it is important to 
 retain the damping term explicitly in 
 $({\mathbf{\Omega}}_0)_{kl} =  \big(\omega_0^2 - \omega_k^2  + \mathrm{i}  z_0 \epsilon^2 \omega_0 \big)\delta_{kl} $ even though it is formally of order $\epsilon^2 $.  It does, however, represent the leading order source of damping in the resonance, which is modified by the final term in \eqref{7=5}, and the combination  turns out to give the correct damping for the effect discussed next. 
 
 \subsection{Resonance conversion by modulation}
We have assumed in Eq.\ \eqref{7=3} that the sum of $\omega_0 $ and the modulation frequency $ \omega_\mathrm{m} $
are close to the modal frequency $\omega_r $, with no restriction on the modulation frequency. 
Now consider the case where the modulation frequency is close to the difference between the modal frequency we have selected, $\omega_r$, and another  lower-valued modal frequency $\omega_t$, specifically
\beq{6=1}
\omega_\mathrm{m} = \omega_r- \omega_t +\frac{\mu}2 \epsilon^2,
\eeq
for some constant $\mu$. 
With this choice of $\omega_\mathrm{m}$ it follows from Eqs.\  \eqref{7=3} and \eqref{6=1} that 
\beq{6=2}
\omega_0 = \omega_t +\frac{(q-\mu)}2 \epsilon^2 . 
\eeq
The drive frequency is close to the resonance frequency $\omega_t$ but the response is clearly of order unity, which is what it would be far from the resonance frequency.  This effect is very different from the high Q resonance of the unmodulated system, as explained next. 

\begin{figure}[ht!] 
\centering
\includegraphics[width=\textwidth]{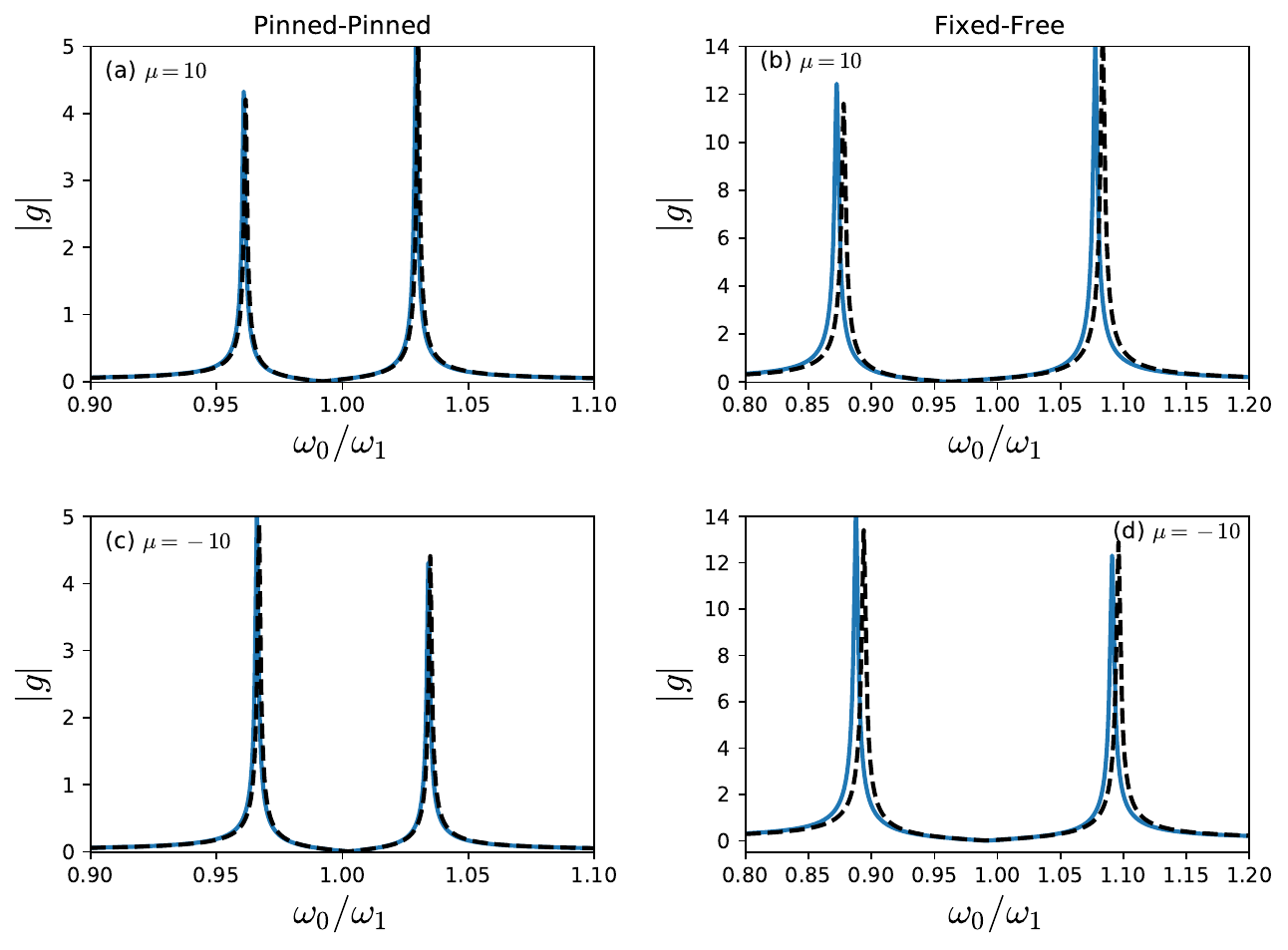}
\caption{The Green's function magnitude of Eq.\ \eqref{9=6} using the precise form of   $\mathbf{\Gamma}_1$ from 
Eq.\ \eqref{1=4} (solid blue curves), as compared with the value from the inner asymptotic expression of Eq.\ \eqref{1=41} (dashed black curves). The coupled modes are $\omega_t= \omega_1$ and $\omega_r= \omega_2$. 
The plots show the magnitude of the Green's function as a function of normalized frequency near the first  resonance frequency of  systems with pinned-pinned (panels (a) and (c)) and fixed-free (panels (b) and (d)) end conditions subject to a modulation frequency $\omega_m $ defined by Eq.\ \eqref{6=1}  with $\mu = \pm 10$. The values of the other parameters are as in Fig.\ \ref{fig1} ($x = 0.43$, $x_s=0.61$, $L=1$, $\epsilon =0.1$,  $Z=0.01$).  }
\label{fig4}
\end{figure}

At the mode of interest, $\omega_0 \approx \omega_t$, Eq.\ \eqref{7=5} can  be written, using Eqs.\  \eqref{7=3} and \eqref{6=1}, as
$
 \big( {\mathbf{\Omega}}_0 -\epsilon^2 \mathbf{\Gamma}_1\big)_{tl}
 = 
 - { K^*_{tr}K_{rl} } /[\omega_r (q +  \mathrm{i}  z_0 ) ] 
 + \mathcal{O}(\epsilon^2)$. 
 Note that the  actual response requires the inverse of the non-diagonal matrix of Eq.\ 
\eqref{7=5}. 
The point is that the modulation causes mode coupling, in this case to effect the  suppression of the resonance at $\omega_0 = \omega_t $ and convert it into two separate quasi-resonances.   A comparison of the exact solution of Sec.~\ref{sec3} with the inner asymptotic expansion of the present Section is given on Fig.\ \ref{fig4}.  
Several things are noteworthy from Fig.\ \ref{fig4}. The appearance of the resonance conversion  effect is independent of the parameter $\mu$ in Eq.\ \eqref{6=1}, indicating that the effect only requires that $\omega_\mathrm{m} \approx  \omega_r- \omega_t $. However, the relative location   and amplitude of the quasi-resonances depends sensitively on  $\mu$. 

\subsubsection{A simplified model}
All of the above numerical examples illustrate how the order unity alteration of the resonance at $\omega_t$ is caused by interaction with the resonance at $\omega_r$, $r>t$.  This suggests that the change in the otherwise diagonal system matrix is from two elements $t$ and $r$. As a first approximation we propose to maintain the diagonal structure of the infinite matrix but retain within the diagonal element the interaction between the two modes of interest. 
This suggests the approximation to Eq.\ \eqref{9=6}
 \beq{9=67}
  g(x,t|x_\mathrm{s}) \approx  \frac{ \psi_t(x)  \psi_t(x_s) }
          { \big( {\mathbf{\Omega}}_0 -\epsilon^2 \mathbf{\Gamma}_1\big)_{tt} }.
\eeq
 The denominator in  Eq.\ \eqref{9=67} follows from Eqs.\ \eqref{6=2} and \eqref{7=5} as
 \bal{7=35}
 \big( {\mathbf{\Omega}}_0 -\epsilon^2 \mathbf{\Gamma}_1\big)_{tt}
 &=  \omega_0^2 - \omega_t^2 +\mathrm{i}\epsilon^2 z_0 \omega_0
 - \frac{ |K_{tr}|^2 } {\omega_r (q +  \mathrm{i}  z_0 ) }  +\ldots 
 \notag\\
 &\approx   \epsilon^2   \omega_t\Big( q +  \mathrm{i}  z_0  -\mu 
 - \frac 1 { \delta^2   (q +  \mathrm{i}  z_0 ) } 
 \Big) 
 \eal
 where $\delta =  { \epsilon \sqrt{ \omega_r \omega_t}}/{ |K_{tr}|}$. 
 Substituting from \eqref{7=35} into \eqref{9=6} gives
  \beq{9=68}
  g(x,t|x_\mathrm{s}) \approx  \frac{ \psi_t(x)  \psi_t(x_s) }
          { \epsilon^2   \omega_t (\alpha + \alpha^{-1})}
\Big( \frac{\alpha}{ q +  \mathrm{i}  z_0  -\frac{\alpha}{\delta} } 
 +  \frac{\alpha^{-1}}{ q +  \mathrm{i}  z_0  +\frac{\alpha^{-1}}{\delta} } \Big)
\eeq
where 
$ \alpha = \frac{\delta \mu}2 + \sqrt{1 + \big(\frac{\delta \mu}2\big)^2}$.
Taking the appropriate asymptotic approximations $\alpha^{\pm 1} = 
1\pm \frac{\delta \mu}2$, and using the original frequency according to \eqref{6=2}, we obtain 
\bal{9=-7}
  g(x,t|x_\mathrm{s}) \approx  \frac 12 \psi_t(x)  \psi_t(x_s) 
\Big( &
\frac{1+ \frac{\epsilon \mu}2  \frac{ \sqrt{ \omega_r \omega_t}}{ |K_{tr}|}}
{ \omega_0^2 - \omega_t^2 +  \big( \frac{\mu}2 + \mathrm{i}z_0\big) \omega_0  \epsilon^2
- \epsilon |K_{tr}| \sqrt{\frac{\omega_t}{\omega_r} } } 
\notag \\ 
+ & 
\frac{1-  \frac{\epsilon \mu}2  \frac{ \sqrt{ \omega_r \omega_t}}{ |K_{tr}|}}
{ \omega_0^2 - \omega_t^2 +  \big( \frac{\mu}2 + \mathrm{i}z_0\big) \omega_0  \epsilon^2
+ \epsilon |K_{tr}| \sqrt{\frac{\omega_t}{\omega_r} } } 
\Big) .
\eal

It is worth revisiting the various parameters in the explicit bi-modal solution of Eq.\  \eqref{9=-7} and the ranges of validity.  The drive frequency $\omega_0$ is assumed to be near the resonance of interest, $\omega_t$, and the modulation frequency $\omega_m$ is assumed to be close to the difference frequency of the two modes $\omega_r>\omega_t $.  The precise form of $\omega_m$ is defined by the tuning parameter $\mu$, see  Eq.\ \eqref{6=1}.  The effect of $\mu$ on the split  resonances  of Eq. \eqref{9=-7} is to shift them higher by $\mathcal{O}(\epsilon^2 )$ and to increase or decrease the magnitudes.   The main cause of the shift in the resonance frequencies are the $\mathcal{O}(\epsilon)$ terms involving the interaction coefficient $K_{tr}$, independent of $\mu$.  This has the effect that the first (second) resonance frequency in  Eq. \eqref{9=-7} is  greater (smaller) than $\omega_t$. 

\begin{figure}[tb!] 
\centering
\includegraphics[width=\textwidth]{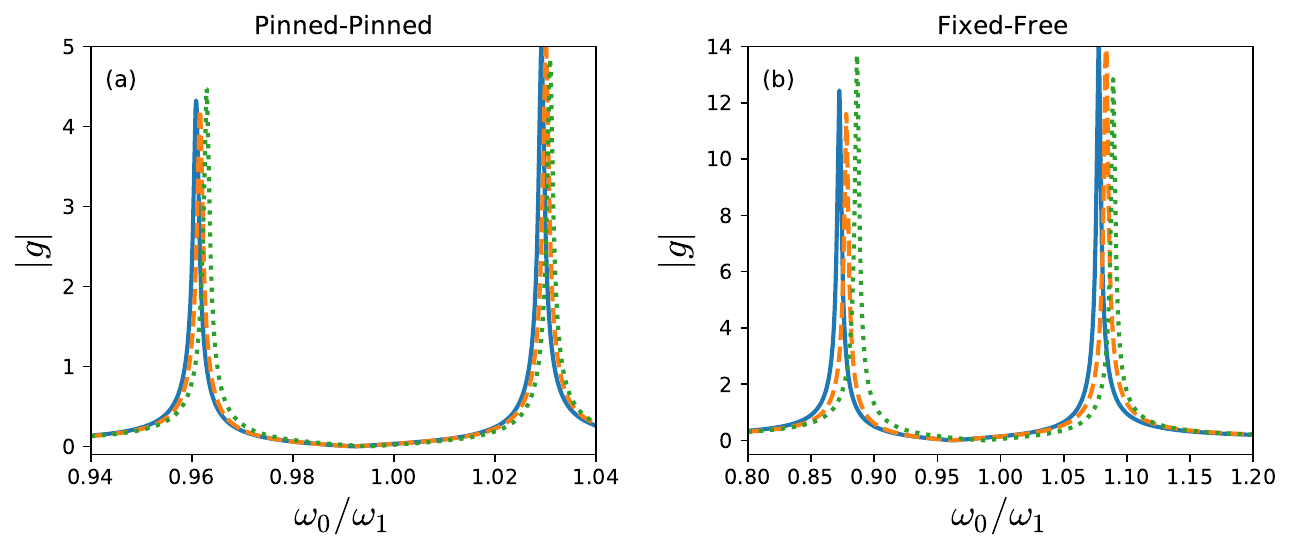}
\caption{The Green's function magnitude of Eq.\ \eqref{9=6} for (a) a pinned-pinned beam and (b) a fixed-free beam using the precise form of   $\mathbf{\Gamma}_1$ from 
Eq.\ \eqref{1=4} (blue curves), the inner asymptotic expression of Eq.\ \eqref{1=41} (dashed orange curves), and the heuristic approximation of Eq.\ \eqref{9=-7} (dotted green curves). The coupled modes are $\omega_t= \omega_1$ and $\omega_r= \omega_2$. 
The  modulation frequency is $\omega_m $ of Eq.\ \eqref{6=1}  with $\mu = 10$. The values of the other parameters are as in Fig.\ \ref{fig1} ($x = 0.43$, $x_s=0.61$, $L=1$, $\epsilon =0.1$,  $Z=0.01$).  }
\label{fig5}
\end{figure}

Figure \ref{fig5} shows how the simplified model compares with the previous ones for the fixed-fixed and free-free examples  considered in Figs.\ 
\ref{fig4}(a) and \ref{fig4}(b).  The two-mode approximation of Eq.\ \eqref{9=-7} provides a good fit to the shifted resonance frequencies, although the amplitudes are not as good a fit as the inner asymptotic approximation of Eq.\ \eqref{1=41}.  This can be attributed to the physically motivated, although somewhat ad-hoc,  diagonal approximation of the global matrix.  Perhaps the main benefit of the approximation \eqref{9=-7} is in its explicit form indicating the role of the modulation frequency tuning parameter $\mu$ and the  coefficient $K_{tr}$ that couples the two interacting modes. 

\section{Conclusions}\label{sec5}
We have developed a Green's function framework to study elastic media with spatiotemporally-modulated material properties. While the focus is on the particular case of flexural waves in a finite beam, the method can be applied to more complex systems where the mode shapes and natural frequencies can be determined numerically, such as with finite element methods. Exact closed-form expressions for the drive frequency component of the Green's function can be found for arbitrary choices of the number of modulated harmonics, $P$, and can be expressed as an asymptotic series in powers of the modulation amplitude $\epsilon$. 
The leading-order corrections to the unmodulated Green’s function near a resonance frequency were obtained via an asymptotic analysis. It was found that the requirements for an order unity effect from the spatiotemporal modulation is that (i) the damping is light in the sense that it is on the order of $\epsilon^2$, and (ii) the modulation frequency is order $\epsilon^2$ close to the difference of two unmodulated modal frequencies. We derived a simple bi-modal solution that is valid for drive frequencies near the lower resonance frequency showing that the modulation splits the lower resonance into two distinct resonances.
The separation of the split resonances is proportional to a parameter $K$ characterizing the coupling between the two modes.
Our asymptotic formulation of the Green's function has identified the scalings necessary to achieve order unity effects from small spatiotemporal modulations in a lightly damped system. This motivates further studies into the unique behavior of spatiotemporally-modulated media, which includes investigating other frequency-mode conversion effects such as nonreciprocity.\\

\noindent \textbf{Acknowledgements}\label{sec6} All authors acknowledge support from the National Science Foundation EFRI Award No.~1641078. B.M.G., S.P.W., and M.R.H. acknowledge additional support from Office of Naval Research under Award N00014-23-1-2660.

\bibliographystyle{RS} 

\end{document}